\documentstyle{l-aa}
\def\degsec#1.#2 {\ifmmode {#1^{\rm o}\hskip-0.42em.
                  \hskip0.10em#2}
         \else {#1$^{\rm o}\hskip-0.42em.\hskip0.10em$#2}
         \fi}

\begin{document}

\thesaurus{09.16.2, 09.04.1}

\title{VLA Measurements of a sample of Planetary Nebulae}

\author{S.R. Pottasch\inst{1}
\and A.A. Zijlstra\inst{2}}

\offprints{S.R  Pottasch}

\institute{Kapteyn Astronomical Institute, P.O. Box 800,
 9700 AV Groningen, the Netherlands
\and
European
Southern Observatory, Karl-Schwarzschild-Strasse 2, D-85748 Garching
bei M\"unchen, Germany}

\date{Received date; accepted date}

\maketitle

\begin{abstract}
We report on new radio measurements of Galactic planetary nebulae,
aimed at resolving the controversies on the reliability of older VLA
flux densities and the suggested deviations from the standard Galactic
extinction law found for planetary nebulae. We show that for faint
($<$10 mJy) objects observed at high angular resolution, previous
determinations are indeed too low. For the bright objects we find no
significant differences. The new values are the most accurate flux
determinations yet for planetary nebulae, reaching 1\% for the
brightest objects in the sample. Based on the new data, we confirm
that there is a systematic difference between the extinction derived
from the radio/H$\beta$ flux ratio and derived from the Balmer
decrement, which led to the suggestion of deviations from the standard
extinction law. However, final confirmation of this has to await the
availibility of more, accurate measurements of the Balmer (and/or
Paschen) lines.

\end{abstract}

\section{Introduction}

In two recent papers (Stasi\'nska et al. 1992; Tylenda et al. 1992) it
has been suggested that the ratio of total to selective extinction,
$R$, is significantly lower than 3, the value corresponding to the
standard extinction law. The authors find that for a sample of
planetary nebulae (PN), the standard extinction law gives a higher
extinction at H$\beta$ from the observed Balmer decrement (effectively
the H$\alpha$/H$\beta$ ratio) than is found from the measured radio to
H$\beta$ flux ratio.  In other words, the observed radio flux is
generally lower than predicted from the H$\beta$
measurements and the Balmer decrement.

The analysis is based on various radio and optical measurements
available in the literature. Most of the radio data comes from two
papers, publishing data obtained with the Very Large Array: Aaquist
and Kwok 1990, and Zijlstra, Pottasch and Bignell 1989.  In the course
of their analysis Tylenda et al.  ``suggest that the radio
measurements of Pottasch's group probably underestimate the radio
fluxes ...  The underestimate becomes more important for fainter
objects, which seems to indicate a sort of instrumental cutoff..'' We
note that the faint objects ($<$10 mJy) all come from the sample of
Zijlstra et al.: Aaquist and Kwok observed brighter sources only.
Two reasons are suggested for the possible underestimation of the
radio flux:
\begin{enumerate}
\item PN may be optically thick at 6 cm, the wavelength most often
observed.
\item VLA measurements are insensitive to extended structures (larger
than 10 times the beam size) which could lead to errors if either a too
small beam was used or an extensive halo exists.
\end{enumerate}

This suggestion can be checked by taking new measurements. In order to
check the first point concerning the optical depth, measurements can
be made at several wavelengths. For practical reasons both 3.6 cm and
6 cm can be measured, the wavelengths at which the system temperature
is lowest. If the nebula is optically thin the flux density at 3.6
cm will be 5\% lower than at 6 cm. If it is completely optically
thick the flux density at 3.6 cm will be considerably higher than at
6 cm, the precise value depending on the geometry and the density
distribution. In order to check the second point (extended
structures), it is possible to increase the beam size by placing the
telescopes closer together (C and D array in the VLA) than was the
case in the earlier measurements. Information on the size of the
object then becomes more uncertain or is completely lost, but the
total flux now includes all possible extended structures. In addition
to checking the earlier radio flux densities, such observations also
allow us to study the first suggestion of the deviating selective
extinction for planetary nebulae.

Because of practical considerations it was impossible to re-do all PN of
interest. Enough time was allotted to re-do 20 PN. A sample was chosen
between $l = 0^\circ$ and $l = 60^\circ$ for which the discrepancy between
the extinction from the Balmer decrement and radio /H$\beta$ ratio was
large. The results of these new measurements are reported here.

\section{Measurements}

The sources were observed at the VLA in September/October 1993, with
the array in the C/D hybrid configuration.  The hybrid configurations
give the best beam shape for southern objects, and the C/D hybrid was
chosen to obtain the largest point-spread function.  The phase
stability is also better at the compact (C,D) configuration, leading
to better calibrated maps.  The observations were therefore optimised
for accurate flux determinations.

All sources were observed for 10 minutes at 6 cm and 4 minutes at 3.6
cm: the integration times can be shorter at 3.6 cm because the
receivers are more sensitive.  Flux calibration was done using 3C 286,
which has a flux density of 7.55 Jy at 6 cm and 5.26 Jy at 3.6 cm
(these values are recent determinations which are 1--1.5\% higher than
previous values). In fact the spectral index of 3C 286 is such that
the flux density differs slightly for both IFs which constitute one
wavelength band, at a separation of 50 MHz. The calibration took this
into account. Calibration and imaging were done using the software
package AIPS. Self calibration was applied in most cases, but was
found not to increase the flux densities significantly for most
sources.  The rms noise in the final maps was 0.1 mJy or slightly
lower at both wavelengths.

A problem was encountered with the AIPS calibration routine in those
cases where the calibrator was only observed once, where the phase
calibration was sometimes faulty. Luckily all affected sources could
be self-calibrated. Flux densities were determined by integrating over
the area of the source. Using the peak flux was not possible because
even at the D-array these PN are slightly resolved. This introduces an
uncertainty especially near the galactic centre, where the many
confusing sources in combination with the poor UV coverage of short
observations cause the field to show residual striping which cannot be
removed. The uncertainty caused by this can be 1 mJy or more, although
having two wavelenghts helps: in general the confusion is less at 3.6
cm. Also, some sources appear to show evidence of extended emission,
where the flux keeps rising if determined over larger areas. In such
cases there is a larger uncertainty in the total flux density, since
the sensitivity of the maps is insufficient to determine the full
extent of the emission, and it is unclear  whether the
extent is indicative of an ionised halo or due to image residuals.

\begin{table*}
\caption[ ]{COMPARISON}
\begin{flushleft}
\begin{tabular}{lllrrcc}
& \multicolumn{1}{c}{Name}&\multicolumn{1}{c}
{ra,dec (1950)}&\multicolumn{2}{c}{S(6
cm) [mJy]}&\multicolumn{1}{c}{S(3.6 cm) [mJy]}&\multicolumn{1}{c}
{Diameter*}\\
& & &\multicolumn{1}{c}{old}& {new} & \multicolumn{1}{c}{new} &
\multicolumn{1}{c}{(arc sec)}\\
& & & & &\\
000.2$-$01.9 & pk0$-$1.5 & 17 50 33.7 $-$29 43 12 & 14.0
& 22.6 & 20.9 & 5\\
000.4$-$02.9 & pk0$-$2.6 & 17 55 07.0 $-$30 00 24 & 5.5
& 7.4 & 7.0 & 7\\
000.7$-$03.7 & pk0$-$3.1 & 17 59 06.6 $-$30 14 28 & 5.0
& 8.7 & 7.8 & 6\\
002.2$-$02.5 & KFL2 & 17 57 50.1 $-$28 16 20 & 2.2 & 2.2
& 2.1 & 5.4\\
002.7$-$04.8 & pk2$-$4.2 & 18 07 54.3 $-$28 59 42 & 24.0
& 28.5 & 26.7 & 9\\
003.9$-$02.3 & pk3$-$2.1 & 18 00 31.9 $-$26 43 42 & 54.0
& 60.5 & 56.6 & 4.4\\
005.7$-$03.6 & KFL13 & 18 09 38.7 $-$25 45 09 & 3.5 & 4.4:
& 4.1: & 14.3\\
005.8$-$06.1 & pk5$-$6.1 & 18 19 46.7 $-$26 50 52 & 3.5
& 20.5 & 18.7 & 5\\
006.4+02.0 & pk6+2.5 & 17 49 40.1 $-$22 21 16 & 59.0
& 60.5 & 59.6 & 7\\
007.5+07.4 & pk7+7.1 & 17 32 14.2 $-$18 32 26 & 3.5 & 7.8
& 8.4 & 6\\
009.8$-$04.6 & pk9$-$4.1 & 18 22 03.6 $-$22 36 35 & 11.0
& 11.9 & 11.5 & 6\\
010.7$-$06.7 & pk10$-$6.2 & 18 31 50.4 $-$22 45 42 & 3.0
& 5: & 4.2 & 7\\
015.9+03.3 & pk15+3.1 & 18 04 40.9 $-$13 29 44 & 98.0 & 99.9
& 97.7 & 4\\
015.6$-$03.0 & pk15$-$3.1 & 18 27 17.3 $-$16 47 23 & 10.0
& -- & 13.2: & 52\\
022.1$-$02.4 & pk22$-$2.1 & 18 37 34.1 $-$10 42 37 & 70.0
& 66.6 & 63.5
& 8\\
024.8$-$02.7 & pk24$-$2.1 & 18 43 51.0 $-$08 31 18 & 3.0
& 12.0 & 11.4 & 5\\
032.1+07.0 & pk32+7.2 &  18 22 13.6 +02 27 45 & 26.0 & 30.6
& 29.3 & 2.7\\
043.1+03.8 & pk43+3.1 & 18 54 12.3 +10 48 10 & 22.0 & 23.0
& 22.8 & 4\\
052.5$-$02.9 & pk52$-$2.2 & 19 36 53.3 +15 49 52 & 44.0
& 45.1 & 42.7 &4.7\\
055.2+02.8 & pk55+2.1 & 19 21 15.1 +21 02 08 & 32.0 & 34.9
& 32.2 & 2.3\\
%& & & mJy & mJy & mJy\\
\end{tabular}

* From Zijlstra et al. (1989), Aaquist and Kwok (1990) and
Kinman et al.
(1988).
\end{flushleft}
\end{table*}

\section{Comparison with previous observations}

Table 1 gives the newly determined flux densities.  The first column
follows the naming convention adopted in the new Catalogue of
Planetary Nebulae (Acker et al. 1992), which we recommend; the second
column gives the name in the old system. Columns 4--6 list the
previous 6 cm flux density and the new determinations at 6 and 3.6 cm.
We note that the 3.6 cm flux is expected to be 5\% below the 6 cm
value if the radio emission is optically thin. This is, in fact, the
case for most of our objects with the exception of PN~006.4+02.0,
007.5+07.4, 015.9+03.3, 032.1+07.0 and 043.1+03.8 which may show
marginal optical depth effects. The flux densities are expected to be
accurate to 1\%, plus an added uncertainty due to image noise, which
will be approximately 1 mJy or less. The fact that the expected
spectral index is found even though the measurements at the two
frequencies are independent of each other, indicates that the accuracy
of the flux densities is good.  These are the most accurate flux
densities for PN yet determined, especially for the brighter objects
where the signal to noise is highest.

However, the comparison of the previous measurement at 6 cm (Zijlstra
et al., 1989; column 5 in Table 1) with the present measurements
(column 6) gives reason for concern. Good agreement (within 10\%) is
obtained in only 50\% of the cases (one of those, PN~032.1+07.0, was
originally observed by Aaquist and Kwok, 1990, rather than Zijlstra et
al.). Marginal agreement (15\% to 50\%) is obtained in 35\% of the
cases, and in 3 cases the differences are even larger. In 2 of these 3
cases, PN~000.2$-$01.9 and 024.8$-$02.7, the quality of the original
image was very poor, as evidenced by a significantly higher noise than
theoretically expected.  In the case of PN~005.8$-$06.1 apparently a
mistake was made in the calibration of the previous measurements. We
finally note that for PN~015.6$-$03.0, the older map was based on a
combination of several observations and is probably better than the
present map due to better UV coverage.

There are 6 sources remaining in the sample with unexplained, large
differences between the new and the old flux densities.  All objects
in our sample with significant discrepancies have three things in
common: they were originally observed in the B-array (i.e.  at high
resolution), are relatively extended ($>5^{\prime\prime}$, a few times
the size of the beam) and are faint (of order 10 mJy or less).  This
suggest that over-resolution of faint sources is the cause of the
problem. For comparison, we note that about 120 of the 300 objects in
the Zijlstra et al. catalogue were observed in the B-array (the others
were observed at lower resolution), and that of those 120 about 30
objects were fainter than 10 mJy and larger than $5^{\prime\prime}$.

Based on this, we suggest that the underestimation of the radio flux
density was caused by a combination of two effects: First, it is very
difficult to obtain an  accurate flux density from a faint, extended
source.  At the time the previous observations were made, accurate
diameters were often not available and errors were made in choosing
the appropriate VLA configuration.  Second, the phase calibration is
more uncertain in the B-array, where significant phase changes can
occur over a few minutes. During long integrations these errors tend
to average out, but for short observations they cause signicant
scattering of flux into the side lobes of the beam.  In general one
use self-calibration to correct for phase errors, but this requires
sufficient signal to noise and is for short observations not possible
below 10 mJy, especially for extended objects.  Both of the discussed
effects affect the same group of sources.

We conclude that of the two explanations for the low radio flux
densities mentioned in the Introduction, the optical depth effects are
not important, but extended structures are. However, the problem is
not the insensitivity of the VLA to highly extended structures (which
begins at larger angular diameters than found the objects observed
here), but rather the decrease in signal to noise for objects a few
times larger than the beam.

Since only about 7\% of the original sample of Zijlstra et al. (1989)
were re-measured, the question arises whether the remainder of the
sample also contains PN whose flux densities have been
underestimated. While nothing can be said with certainty, it seems
likely that the worst cases were part of the sample which were
re-measured. The reason for this is that this sample was chosen from
among those cases where there was poor agreement with the optical
measurements. We therefore expect that the accuracy of the remainder
of the original sample is better.  As mentioned above, about 30
objects (including the presently observed objects) fall in the 'danger
zone', with diameter larger than $5^{\prime\prime}$, flux density less
than 10 mJy and originally observed in the B-array. All these should
be re-observed.  The catalogue of Aaquist and Kwok (1990) contains
brighter sources and they also used longer integration times: we therefore
expect that their catalogue will be less affected by these problems.

We finally note that the encountered problems appear to be inherent to
snapshot observations, and that in the extended VLA configurations,
longer observations are desirable to obtain accurate flux densities
for extended objects.

\section{Comparison of extinction}

It may now be asked whether the new radio observations alter in any way
the conclusions of Stasi\'nska et al. (1992) and Tylenda et al. (1992),
discussed in the Introduction.

In Fig.\ 1 the value of $C(= \log[{\rm H}\beta_{\rm exp}/ {\rm
H}\beta_{\rm obs}])$, where H$\beta_{\rm obs}$ is the observed
H$\beta$ flux and H$\beta_{\rm exp}$ is the value predicted from the
observed radio data, is plotted as ordinate. The conversion from radio
data to H$\beta_{\rm exp}$ is made assuming the electron temperature
($T_{\rm e}$) of the gas is 10$^{4}$ K and the He/H ratio is 0.1 with
all the helium singly ionized. These assumptions could introduce a
20\% error in H$\beta_{\rm exp}$, but it is unlikely to be more than
this. On the abscissa, the value of $C$ is plotted, derived by
assuming a standard extinction law (e.g. Seaton, 1979) and adjusting
the amount of extinction so that the observed H$\alpha$/H$\beta$ ratio
is reduced to the theoretically expected value at $T_{\rm e}=10^{4}$ K
of 2.85 (Brocklehurst, 1971).

The open circles on Fig.\ 1 are measurements from the survey of Acker
and Stenholm (Acker et al. 1991; Tylenda et al. 1992). All nebulae
which we have measured have data from this source. The quality of
these data varies, depending on the size and brightness of the
nebula. For example, the H$\beta$ flux was measured with a square
diaphragm $4^{\prime\prime}$ on a side. For larger nebulae a
correction factor was applied, but this can be uncertain. The
triangles in the figure use the same Balmer decrement but H$\beta$
fluxes which are more accurate, taken from Shaw and Kaler (1989),
Webster (1983) and O'Dell (1963). A line connecting the circle and the
triangle indicates that the same nebula is involved.  Unfortunately
only 6 PN have more accurate H$\beta$ flux measurements, but they
indicate that the flux measurements of Acker et al. (1991) are accurate
to the limits given by these authors\footnote{Kinman et al.  (1988)
also have measured H$\beta$ fluxes in about half of these
objects. They are sometimes quite different from the measurements of
Acker et al. and we have chosen to ignore them}.

The crosses in Fig.\ 1 indicate the few other measurements of Balmer
decrements available in the literature (Shaw and Kaler 1989, Kinman et
al. 1988, Aller and Keyes 1987, de Freitas Pacheco et al. 1992, Ratag
1991). As can be seen from the figure, in half the cases good
agreement is obtained with the measurements of Tylenda et al. while in
the other half substantial differences are found. It is clear that
more, careful measurements are desirable.

\begin{figure}
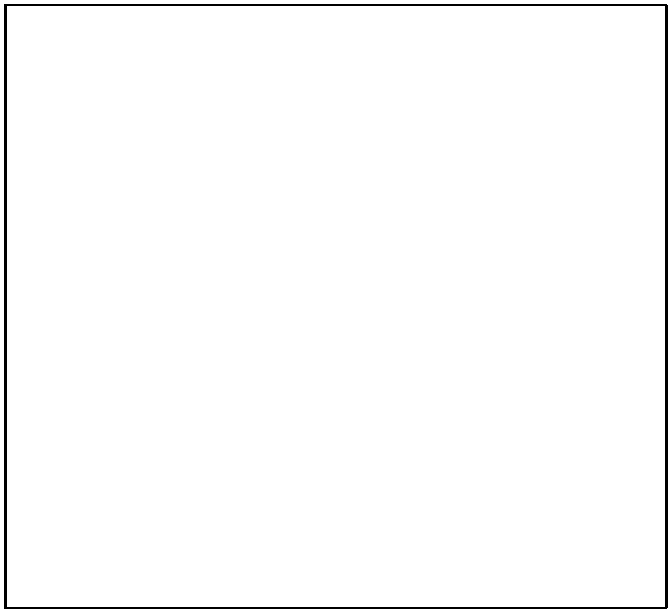

\picplace{8 cm}
\caption[ ] {The extinction coefficient $c$ derived from the Balmer
decrement versus the value derived from the radio/H$\beta$ flux ratio,
for the PN discussed in this paper.  Open circles indicate
measurements from Acker et al. (1991) and Tylenda et al. (1992).  The
triangles present H$\beta$ fluxes from different papers, while the
crosses present different measurements of the Balmer decrements (for
references see text). The connecting lines indicate that the same nebula is
involved.}
\end{figure}

As can be seen in Fig.\ 1, a straight line which best represents the
observed circles is less steep than the $45^\circ$ line shown in the
diagram. This is also the conclusion of Stasi\'nska et al. and Tylenda
et al. It indicates that the new radio measurements do little to
affect this conclusion. This fact can be interpreted in two ways;
either (1) the total to selective extinction is lower than 3.1, the
standard extinction law, or (2) the observational errors, especially
of the Balmer decrement, are still too uncertain to conclude that the
points do not fit the $45^\circ$ line. On the basis of the nebulae
plotted in Fig.\ 1, the latter alternative appears quite real.

We note that a similar discrepancy, with radio flux density too low
compared to the H$\beta$ flux, has recently been found for two HH
objects: HH 32A and HH 1-2 (Anglada et al. 1992).  Anglada et
al. suggest as explanation the presence of significant shock
excitation of the H$\alpha$ line, although they also note that the
shock velocities in these objects appear too high for this mechanism
to work. Since their observations were done in the VLA
D-configuration, extended emission is not expected to contribute to
the deficit. This provides still another possible interpretation of the
discrepant measurements.

Whichever of these alternatives is correct however, one thing is
certain: use of the value of $C$ obtained from the Balmer decrement
measures of Tylenda et al. (1992) together with the standard
extinction curve, will overestimate the extinction in the visual or
ultraviolet.  Thus this method should not be used, e.g. to correct the
visual magnitudes of the central stars. In this case the extinction
correction should be made from the radio/H$\beta$ extinction value.

\end{document}